\documentclass[aps, reprint, prl, twocolumn, superscriptaddress]{revtex4-1}

\usepackage{graphicx}
\usepackage{dcolumn}
\usepackage{bm}
\usepackage{amsmath}
\usepackage{amssymb}
\usepackage{upgreek}
\usepackage{braket}
\usepackage{color}
\usepackage{textcomp}
\usepackage{placeins}
\usepackage{nicefrac}
\usepackage{ bbold }

\begin{document}

\title{Anti-Localization in Oxides: Effective Spin-$\boldsymbol{\nicefrac{3}{2}}$ Model}

\author{Patrick Seiler}
\email{patrick.seiler@physik.uni-augsburg.de} 
\affiliation{Center for Electronic Correlations and Magnetism, EP VI, Institute of Physics, University of Augsburg, 86135 Augsburg, Germany}

\author{Elias Lettl}
\affiliation{Center for Electronic Correlations and Magnetism, EP VI, Institute of Physics, University of Augsburg, 86135 Augsburg, Germany}

\author{Daniel Braak}
\affiliation{Center for Electronic Correlations and Magnetism, EP VI, Institute of Physics, University of Augsburg, 86135 Augsburg, Germany}
\affiliation{Max Planck Institute for Solid State Research, Heisenbergstra{\ss}e 1, 70569 Stuttgart, Germany}

\author{Thilo Kopp}
\affiliation{Center for Electronic Correlations and Magnetism, EP VI, Institute of Physics, University of Augsburg, 86135 Augsburg, Germany}

\date{\today}


\begin{abstract}

Weak anti-localization offers an experimental tool to address spin--orbit coupling of two-dimensional oxide surfaces and interfaces via magneto-transport. To overcome the shortcomings of the formulation for single-band spin-$\nicefrac{1}{2}$ electrons, we consider an effective three-band model that allows a decomposition into a pseudo-spin representation $\nicefrac{1}{2} \oplus \nicefrac{3}{2}$. Whereas the well-established spin-$\nicefrac{1}{2}$ transport signature results from the singlet and triplet sectors in the Cooperon equation, a new structure originates from the quintet and septet sectors generated by the spin $\nicefrac{3}{2} \otimes \nicefrac{3}{2}$ representation.

\end{abstract}



\maketitle


A disordered two-dimensional electronic system is marginal with respect to electron propagation, residing at a metal--insulator transition.
Quantum corrections to conductivity turn the system into a metallic or an insulating state. 
On the one hand, self-interference of an electron wave in the presence of considerable disorder results in Anderson localization~\cite{PhysRev.109.1492, Wegner1976, PhysRevB.22.4666, PhysRevLett.45.842, RevModPhys.80.1355}. Even for arbitrarily weak disorder, the effect of weak localization~\cite{Hikami01021980, PhysRevB.22.5142, RevModPhys.57.287} results in an insulating state in two dimensions~\cite{PhysRevLett.42.673, Wegner1979}. 
On the other hand, the relativistic effect of spin--orbit coupling (SOC) creates a metallic state in the two-dimensional system~\cite{Hikami01021980}. The underlying symplectic symmetry results in a metal--insulator transition~\cite{Wegner1989663} as well as weak anti-localization (WAL). Both, weak localization and WAL originate from elastic scattering processes at impurities that produce a singular contribution for momentum transfer at  twice the Fermi momentum (Cooperon).

These quantum corrections are quite distinctive in magneto-transport measurements and, in case of noticable SOC, cause a WAL signature in form of a minimum in the perpendicular field magneto-conductivity~\cite{Bergmann19841, PhysRevB.87.245121, PhysRevLett.108.206601, PhysRevB.97.075136, 0034-4885-81-3-036503}. Hikami, Larkin and Nagaoka~\cite{Hikami01021980} have derived the first order WAL quantum correction based on the Elliott--Yafet~\cite{PhysRev.96.266, Yafet1983287} spin relaxation mechanism, where a spin flip may occur during the scattering process. In this case, it has been concluded that the specific shape of the Fermi surface or multi-band effects only play a minor role, which is captured by adaptation of fitting parameters~\cite{PhysRevB.32.3522}. However, as already stated in the original Hikami--Larkin--Nagaoka paper~\cite{Hikami01021980}, the Elliott--Yafet spin relaxation cannot result in a WAL correction for a truly two-dimensional electron system, because the corresponding contribution is non-zero only for sizable scattering processes perpendicular to the two-dimensional plane. 

Later, Iordanskii, Lyanda-Geller, and Pikus introduced a WAL theory that describes the in-plane SOC of a two-dimensional system~\cite{JETPLett60206, PhysRevB.51.16928, PhysRevB.53.3912} by consideration of the D'yakonov--Perel spin relaxation~\cite{dyakonov1971spin}. This kind of spin relaxation typically occurs in electronic systems with broken inversion symmetry and describes a decay of the spin expectation value according to spin precession in a random effective field. The D'yakonov--Perel spin relaxation is intrinsically connected to the Rashba and Dresselhaus effects that manifest themselves in semiconductor quantum wells as well as oxide interfaces and heterostructures. In Iordanskii--Lyanda-Geller--Pikus theory, both the shape of the Fermi surface and the spin structure in momentum space have a major effect on the magneto-conductivity, as opposed to the earlier theory. Interestingly, in the particular case that the spin expectation value, traced along the Fermi surface, can be described by a threefold winding number $N_\text{S}=3$, the formula of this approach reproduces the analytical structure of the formula of the Hikami--Larkin--Nagaoka calculation. However, in the case of a non-zero single winding contribution $N_\text{S}=1$, which is typical for Rashba and Dresselhaus effects, the results deviate considerably from Hikami--Larkin--Nagaoka theory~\cite{JETPLett60206}, caused by transitions between dense Landau levels in a magnetic field. Correspondingly, the specifics of the Fermi surface have a strong impact on the magneto-conductivity for two-dimensional electron systems with SOC~\cite{PhysRevB.97.075136}.

The electronic interface in the LaAlO$_3$/SrTiO$_3$ heterostructure~\cite{Ohmoto2004} has become paradigmatic for a strongly confined electron system, with the band filling tuned via gate voltage~\cite{Thiel29092006, :/content/aip/journal/apl/100/5/10.1063/1.3682102, :/content/aip/journal/aplmater/3/6/10.1063/1.4921068}, and a sizable Rashba-like~\cite{0022-3719-17-33-015, Winkler} spin--orbit effect. The corresponding strength of the SOC seems to be strongly dependent on the electron density in the system and reaches a maximum near the Lifshitz transition, which is related to the avoided Ti 3d t$_{2\text{g}}$ band crossings~\cite{PhysRevLett.104.126803, Ilani2012}. The control of SOC at oxide interfaces is considered to be a great opportunity for spintronics in future electronic devices~\cite{Cen1026, Mannhart1607, :/content/aip/journal/apl/100/5/10.1063/1.3682102}. 
For this LaAlO$_3$/SrTiO$_3$ interface, and for transition metal oxide materials in general, the hitherto existing WAL description by single-band electronic states is a questionable simplification.  
It is to be expected that the elastic impurity scattering for $N_\text{S}=1$ in the Cooperon involves interband scattering jointly with the transitions between Landau levels, which has a yet to be determined impact on the WAL in these systems.
To our knowledge, the required multi-band analysis of WAL has not been achieved. 

In this article, we devise such a more realistic WAL starting with a three-band Hamiltonian, suitable for oxide interfaces and heterostructures. We consider an effective model that allows for a group-theoretical analysis, reducing to a spin--orbit coupled quartet with spin quantum number~$\nicefrac{3}{2}$ as well as a doublet with spin quantum number~$\nicefrac{1}{2}$. It turns out that the coupling of Landau levels within the spin multiplets of the $\nicefrac{3}{2} \otimes \nicefrac{3}{2}$ representations in the Cooperon channels is the key mechanism for the resulting magneto-conductivity behavior.

\begin{figure}
    \includegraphics[width=0.95\columnwidth]{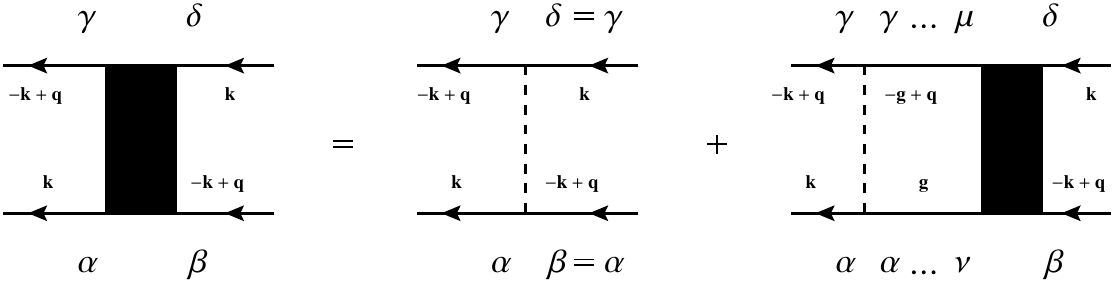}
    \caption{ \label{Fig:Pic1} \emph{Dyson equation for the Cooperon.} The Cooperon is characterized by momentum transfer of $\Delta \mathbf{k} = 2 \mathbf{k}_\text{F}$ through the maximally crossed diagrams. Spin relaxation occurs during propagation due to D'yakonov--Perel spin relaxation.}
\end{figure}

The quantum correction to the conductivity is given by~\cite{PhysRevB.22.5142}
\begin{equation}
    \delta \sigma(\omega \rightarrow 0) = -2 e^2 N_\text{F} D \tau_0^2 \sum\limits_{\alpha \beta} \int\limits_{\mathbf{q}} C_{\alpha \beta \beta \alpha} (\mathbf{q} ),
\end{equation}
where $N_\text{F}$ is the Fermi level density of states, $D=v_\mathrm{F}^2 \tau_1 /2$ the diffusion constant, $\tau_1$ the first harmonic or transport time and $\tau_0$ the elastic scattering time. The Cooperon components in electron spin space, $C_{\alpha \beta \beta \alpha} (\mathbf{q})$, are found by solving the momentum dependent Dyson equation, see Fig.~\ref{Fig:Pic1}.
Maximally crossed diagrams describe a momentum transfer, where transitions with $\Delta \mathbf{k} = 2 \mathbf{k}_\text{F}$ are dominant due to the Cooperon pole structure, and spin relaxation occurs during propagation due to D'yakonov--Perel spin relaxation. 

The corresponding Green's functions relate to the single-band Hamiltonian
\begin{equation}
    \mathcal{H}_\text{single} = \frac{\hbar^2 \mathbf{k}^2}{2 m} \otimes \mathbb{1}_{2 \times 2} + \hbar \boldsymbol{\upsigma} \cdot \boldsymbol{\mathrm{\Omega}} (\mathbf{k}) 
    \label{eq:Hamfourbands}
\end{equation}
where $\mathbb{1}_{2 \times 2}$ is the unity matrix in spin space, $\boldsymbol{\upsigma} = ( \sigma_\text{x}, \sigma_\text{y} )^\text{T}$ is the two-dimensional vector of Pauli spin matrices within the plane, and $\boldsymbol{\mathrm{\Omega}}(\mathbf{k}) = - \boldsymbol{\mathrm{\Omega}}(-\mathbf{k})$ is the spin--orbit vector potential.
Specifically, an expansion into first and third harmonic of the vector field for Dresselhaus SOC is
\begin{equation}
    \boldsymbol\Omega (\textbf{k}) = \Omega_1 (k) \begin{pmatrix}
         -\cos \phi \\
         \phantom{-}\sin \phi
    \end{pmatrix}
    + \Omega_3 (k) \begin{pmatrix} \cos \left( 3 \phi \right) \\ \sin \left( 3 \phi \right) \end{pmatrix},
    \label{eq:genvec}
\end{equation}
where $\tan \phi = k_\mathrm{y}/k_\mathrm{x}$ refers to the angle in momentum space and $\Omega_1$ and $\Omega_3$ measure the strength of the vector potential causing a $N_\text{S}=1$ and $N_\text{S}=3$ spin winding, respectively. For (linear) Rashba SOC, $\boldsymbol\Omega (\textbf{k}) = \Omega_1 (k) \cdot ( \sin \phi, -\cos \phi)^{\rm T}$. 

The Dyson equation (see Fig.~\ref{Fig:Pic1}) is solved by summation of ladder diagrams in the particle--particle channel in presence of a magnetic field $B$~\cite{JETPLett60206, PhysRevB.51.16928, PhysRevB.53.3912}.
The characteristic magnetic fields $B_\text{so/i}^{(\prime)}$ are given by 
\begin{equation}
    B_\text{so/i}^{(\prime)} = \frac{\hbar}{ 4 e D \tau_\text{so/i}^{(\prime)} },
\end{equation}
where $\tau_\text{i}$ is the relaxation time for inelastic scattering processes, and $\tau_\text{so}^{(\prime)}$ are the time scales related to the D'yakonov--Perel spin relaxation~\cite{dyakonov1972spin}, defined by~\cite{JETPLett60206}
\begin{align}
    \frac{1}{\tau_\mathrm{so}}        &= 2 \left( \Omega_1^2 \tau_1 + \Omega_3^2 \tau_3 \right) \\[0.2cm]
    \frac{1}{\tau_\mathrm{so}^\prime} &= 2 \Omega_1^2 \tau_1, 
    \label{eq:relax}
\end{align}
where $\Omega_{1 (3)} = \Omega_{1 (3)}(k=k_\text{F})$ are given by their Fermi surface values and $\tau_{1(3)}$ are the relaxation times of the first (third) harmonic of the Cooperon~\cite{JETPLett60206}. Note that the result for the magneto-conductivity itself does not distinguish between the Rashba and the Dresselhaus vector structure, as the result involves the spin--orbit scattering times only. 

In order to determine the WAL contribution analytically for a multi-band case, it is unavoidable to implement severe restrictions. 
Explicitly, we investigate the three-band case with degenerate (parabolic) bands in the vicinity of the $\Gamma$-point. We will comment on these restrictions below. 
The form of the atomic SOC and the inversion symmetry breaking terms, as presented below, are taken in the spirit of the effective three-band model for LaAlO$_3$/SrTiO$_3$ interfaces~\cite{PhysRevB.87.161102}, in which the Ti 3d t$_{2\text{g}}$ bands constitute the local basis. 
The atomic spin--orbit coupling in the local basis is given by~\cite{PhysRevB.87.161102}
\begin{equation}
    \mathcal{H}_\text{aso} = \Delta_\text{aso}
    \begin{pmatrix}
        0 & i \sigma_{\text{z}} & -i \sigma_{\text{y}} \\
        -i \sigma_{\text{z}} & 0 & i \sigma_{\text{x}} \\
        i \sigma_{\text{y}} & - i \sigma_{\text{x}} & 0
    \end{pmatrix}.
    \label{eq:haso}
\end{equation} 
The eigenstates of this Hamiltonian group into a higher energy doublet and a lower energy quartet, separated by $3 \Delta_\text{aso}$~\cite{Callaway}. In addition, we allow a small intermixing due to the asymmetry of the interface, with strength
$\Delta_\text{m}(k)$ that is linear in $k$ for small $k$~\cite{PhysRevB.87.161102},
\begin{equation}
    \mathcal{H}_\text{m} = \Delta_\text{m}(k)
    \begin{pmatrix}
        0 & 0 & i \cos(\phi) \\
        0 & 0 & i \sin(\phi) \\
        - i \cos(\phi) & -i \sin(\phi) & 0
    \end{pmatrix} \otimes \mathbb{1}_{2 \times 2}.
    \label{eq:hm}
\end{equation} 
The combination results in
\begin{equation}
    \begin{split}
    &\mathcal{H}_\text{aso} + \mathcal{H}_\text{m} = \\[0.2cm]
    &= \begin{pmatrix}
        2 \left[ \Delta_\text{aso} \otimes \mathbb{1}_{2 \times 2} + \hbar \boldsymbol{\upsigma} \cdot \boldsymbol{\Omega} \right] & \Delta_\text{m}\, A^\dagger \\[0.3cm]
        \Delta_\text{m}\, A & -  \Delta_\text{aso} \otimes \mathbb{1}_{4 \times 4} + \hbar \mathbf{S} \cdot \boldsymbol{\Omega} 
    \end{pmatrix},
    \end{split}
    \label{eq:h66}
\end{equation} 
where 
\begin{equation}
    \boldsymbol{\Omega} = - \frac{\Delta_\text{m}}{3 \hbar } 
    \begin{pmatrix}
        - \cos \phi  \\
        \phantom{-} \sin \phi 
    \end{pmatrix},
\end{equation}
and $\mathbf{S}/2$ is the two-dimensional vector of matrices that form a spin-$\nicefrac{3}{2}$ representation of the angular momentum algebra. $A$ is a $4\times 2$-matrix,
where non-zero terms depend on momentum only in terms of the angle $\phi$, but not on the absolute value.

As we assume $\Delta_\text{m} \ll \Delta_\text{aso}$, the coupling between doublet and quartet is small and we can project to the quartet and doublet subspaces in lowest
order. The effective SOC in the quartet subspace is explicitly given by
\begin{equation}
    \mathcal{H}_\text{SOC}^\text{quartet} = \hbar \mathbf{S} \cdot \boldsymbol{\Omega} = 
    \begin{pmatrix}
        0 & \sqrt{3} \omega & 0 & 0 \\
        \sqrt{3} \omega^\ast & 0 & 2 \omega & 0 \\
        0 & 2 \omega^\ast & 0 & \sqrt{3} \omega \\
        0 & 0 & \sqrt{3} \omega^\ast & 0
    \end{pmatrix},
    \label{eq:h32}
\end{equation} 
where $\omega = \Omega_\text{x} - i \Omega_\text{y} $. Evidently, Eq.~\eqref{eq:h32} represents an effective
momentum dependent spin--orbit coupling~\cite{Winkler}. Note that the spin--orbit coupling for spin-$\nicefrac{3}{2}$ still belongs to the symplectic symmetry class~\cite{PhysRevB.81.045104}.

In the following, we discuss only the case of $N_\text{S}=1$ spin winding (i.e. $\Omega_3=0$) and refer to the detailed calculations and results for general mixed $N_\text{S}=3$/$N_\text{S}=1$ spin winding in the supplement~\cite{Supp}.
We derive the WAL correction for Hamiltonian Eq.~\eqref{eq:h32} in analogy to the Iordanskii--Lyanda-Geller--Pikus calculation~\cite{JETPLett60206, PhysRevB.51.16928, PhysRevB.53.3912}.
Note that we focus on the low-energy quartet section of the Hilbert space, as WAL for the high-energy doublet subspace is described by the original theory.

\begin{figure}
    \includegraphics[width=0.95\columnwidth]{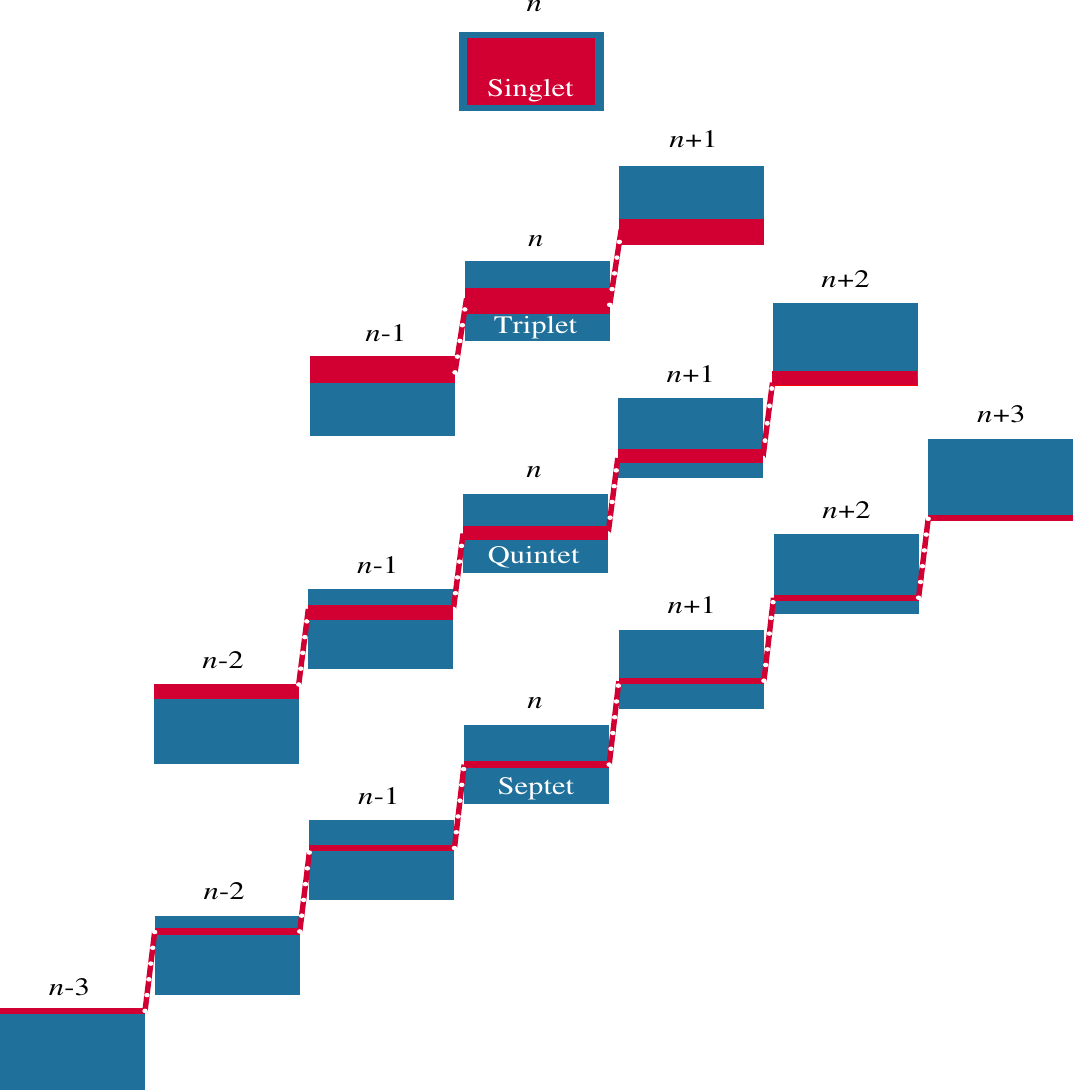}
    \caption{ \label{Fig:Pic2} \emph{Coupling of effective Landau levels in the Cooperon equation.} A magnetic field introduces Landau level quantization. For the singlet Cooperon channel, there is only one Landau level. For the triplet case, each Landau level splits into three sublevels, and accordingly also in the higher spin channels. A single spin winding contribution with $N_\text{S}=1$ in the Hamiltonian couples sublevels from near Landau levels.}
\end{figure}

In the Cooperon equation, we find the product of Green's functions in generalized spin space,
\begin{equation}
    \frac{\hbar}{\xi_{\mathbf{k}} + \hbar \mathbf{S}\cdot \boldsymbol{\Omega} \left( \mathbf{k} \right) - \frac{i \hbar}{2 \tau} } \times
    \frac{\hbar}{\xi_{-\mathbf{k}+\mathbf{q}} + \hbar \mathbf{R}\cdot \boldsymbol{\Omega} \left( -\mathbf{k} + \mathbf{q} \right) + \frac{i \hbar}{2 \tau} }
    \label{eq:prodG}
\end{equation}
where $\tau^{-1} = \tau_{\text{i}}^{-1} + \tau_0^{-1}$, $\xi_{\mathbf{k}}$ is the kinetic energy, and $\nicefrac{\mathbf{R}}{2}$ constitutes the spin-$\nicefrac{3}{2}$ algebra of the second Green's function. 
The expression Eq.~\eqref{eq:prodG} is expanded in vicinity of the Fermi surface. The combination of two spin-$\nicefrac{3}{2}$ particles decomposes into the spin channels
\begin{equation}
    \frac{3}{2} \otimes \frac{3}{2} = 0 \oplus 1 \oplus 2 \oplus 3,
    \label{eq:decomp}
\end{equation}
that is a spin singlet, triplet, quintet, and septet channel.
The Cooperon equation can be solved by a summation of the inverse eigenvalues of the Cooperon inverse, $\mathcal{L}$.
In a magnetic field $B$, the eigenvalues of the distinct spin channels are regrouped in Landau levels, see Fig.~\ref{Fig:Pic2} for a visualization.
The magneto-conductivity is given by
\begin{equation}
    \begin{split}
    \Delta \sigma &= \frac{e^2 \kappa}{2 \pi h} \sum\limits_{n=0}^{\hspace{0.3cm}(\kappa \tau_1)^{-1}} \left( -\frac{1}{E^{(0)}(n)} \hspace{0.2cm} \right. + \sum\limits_{ \tilde{m} = 0, \pm 1 } \hspace{0.1cm} \frac{1}{E^{(1)}_{\tilde{m}} (n)} \hspace{0.2cm}  \\  &-\sum\limits_{ \tilde{m} = 0, \pm 1, \pm 2 } \hspace{0.1cm} \frac{1}{E^{(2)}_{\tilde{m}} (n)} \hspace{0.2cm} + \left. \sum\limits_{ \tilde{m} = 0, \pm 1, \pm 2, \pm 3 } \hspace{0.1cm} \frac{1}{E^{(3)}_{\tilde{m}} (n)} \right),
    \end{split}
    \label{eq:MCsuminverse}
\end{equation}
where $\kappa = 4 e B D \tau_1 / \hbar$ and $\tilde{m}$ is the internal quantum number of the decomposition Eq.~\eqref{eq:decomp}.
Note that the singlet and quintet channels contribute with a negative sign, whereas the triplet and septet contribute with a positive sign in Eq.~\eqref{eq:MCsuminverse}.

In the case of exclusive $N_\text{S}=3$ winding, i.e.\ $\Omega_1=0$ and consequently $B_\text{so}^\prime=0$, Landau levels are not intermixed. The sum over inverse eigenvalues can then be calculated directly. In the supplement, we provide an analytic formula for this case, using digamma functions~\cite{Supp}.

However, for a finite $N_\text{S}=1$ winding contribution, and $\Omega_1 \neq0$, 
the first harmonic of the spin--orbit vector potential results in a pairing of Landau level transitions (described by Landau ladder operators $a$, $a^\dagger$) with total spin change (denoted by total spin ladder operators $J_+$, $J_-$ as defined in the supplement~\cite{Supp}) in the $\mathcal{L}$ operator of the kind 
$\sqrt{ B_\text{so}^\prime / B  } (J_+ a + J_- a^\dagger$),
as already observed in the original spin-$\nicefrac{1}{2}$ evaluation.
For the single-band spin-$\nicefrac{1}{2}$ case, in the triplet channel the spin projection changes jointly with a transition to the next Landau level. This is similar for the pseudospin-$\nicefrac{3}{2}$ case, although here quintet and septet ladders are also involved (see Fig.~\ref{Fig:Pic2}). Moreover, the change in the spin projection is a combined spin--orbital transition, that is, a band transition is also implied. 
This mixing of Landau levels produces for Dresselhaus SOC a matrix structure known for the Tavis--Cummings model, which is a generalization of the Jaynes--Cummings model~\cite{Allen:2019274, Supp}. For Rashba SOC it is the anti-Jaynes--Cummings model~\cite{PhysRevA.97.023624} because there a stepwise increase/decrease of spin goes along with a stepwise increase/decrease in the Landau level $n$.
For either, Dresselhaus or Rashba SOC, the inverse eigenvalues can be treated analytically but embrace a large number of terms in the magneto-conductivity in Eq.~\eqref{eq:MCsuminverse}. 
Detailed expressions can be found in the supplement~\cite{Supp}.

\begin{figure}
    \includegraphics[width=0.95\columnwidth]{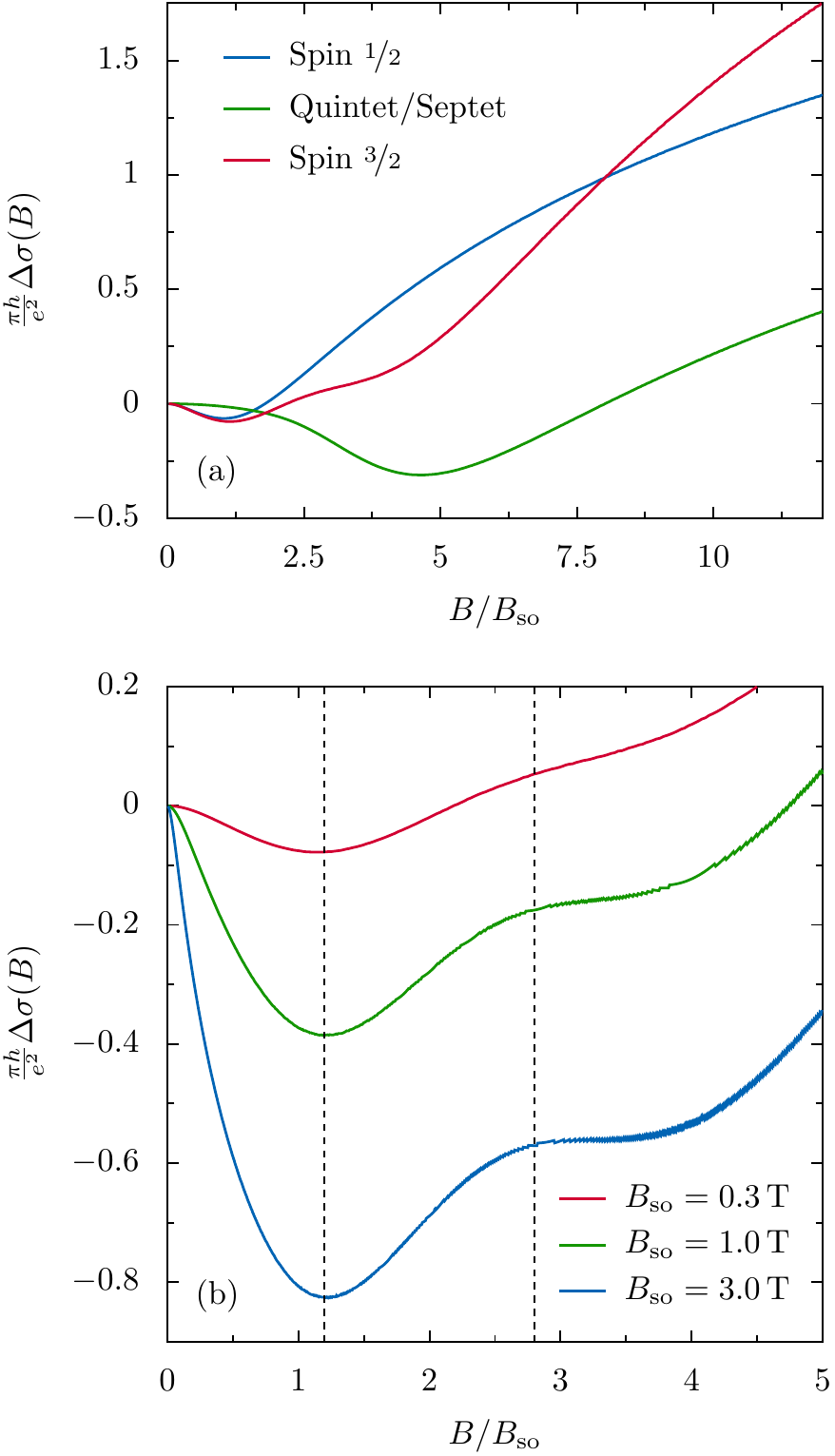}
    \caption{ \label{Fig:Pic3} \emph{Additional structure in the magneto-conductivity of three-band weak anti-localization.} (a) $B_\text{so}=B_\text{so}^\prime=0.3$\,T. The combination of quintet and septet contribution (green line) of the spin-$\nicefrac{3}{2}$ Cooperon shows an additional minimum. 
The full spin-$\nicefrac{3}{2}$ magneto-conductivitiy (red line) is composed of the spin-$\nicefrac{1}{2}$ response (blue line) and the quintet/septet contribution, creating the shoulder-like structure. 
(b) Spin-$\nicefrac{3}{2}$ magneto-conductivity response for different spin--orbit fields $B_\text{so}=B_\text{so}^\prime$. 
The position of the local minima in the magnetic field dependency scale with the spin--orbit field.}
\end{figure}

The magneto-conductivity of this effective three-band model is qualitatively different from that of the single-band spin-$\nicefrac{1}{2}$ case: we identify an additional shoulder-like structure in the spin-$\nicefrac{3}{2}$ section that is caused by a minimum 
in the combined quintet and septet contribution (see Fig.~\ref{Fig:Pic3}a, green line). The spin-$\nicefrac{1}{2}$ (blue line) and spin-$\nicefrac{3}{2}$ (red line) cases can be well distinguished. 

The winding number $N_\text{S}$ is  a topological invariant, having consequences for transport characteristics. It appears that the case $N_\text{S}=1$ has a distinct signature in magneto-transport: it couples multiple Landau levels. 
A measurable indication of this coupling is, in the single-band case, a minimum at  $B\sim B_\text{so}$ in the magneto-conductivity that results from the joint contribution of the singlet and triplet channels. For the three-band scenario with $N_\text{S}=1$, when it can be mapped to a spin-$\nicefrac{3}{2}$ case, a minimum is generated from the joint contribution of the singlet and triplet channels and a further minimum from the quintet and septet channels, both of which scale with $B\sim B_\text{so}$ but with different prefactors (see Fig.~\ref{Fig:Pic3}b).
It is clear that in our effective model spin flips and band transitions are controlled by the same energy scale---therefore, a new scale is not expected.

On the other hand, electronic states with $N_\text{S} >1$ do not introduce a coupling of sublevels from different Landau levels. Consequently,  
a clear shoulder-like structure
is absent for $N_\text{S}>1$. 
This finding is certainly a helpful rule as $N_\text{S}$ is not easily measured by other means. 
 
The presented model is built on symmetry assumptions that allow to decompose the six-dimensional representation into a doublet (spin-$\nicefrac{1}{2}$) and a quartet (spin-$\nicefrac{3}{2}$) irreducible representation and made the analytical analysis possible. Symmetry breaking terms produce a mixing of the higher energy spin-$\nicefrac{1}{2}$ states with the spin-$\nicefrac{3}{2}$ states, split by $3\Delta_\text{aso}$. Explicitly, the orbital intermixing of Eq.~(\ref{eq:hm}) also generates a mixing of
doublet and quartet blocks which may be neglected when $\Delta_\text{m}/\Delta_\text{aso} \ll 1$. Similarly, a band gap at the $\Gamma$-point mixes the blocks if this gap is not small with respect to $3\Delta_\text{aso}$. Material specific models will certainly go beyond the analytical technique shown here. A full numerical treatment of the spin-$\nicefrac{3}{2}$ Cooperon equation might become available in the near future~\cite{PhysRevE.95.023309}.

In conclusion, we formulated analytically weak anti-localization for a three-band model, suitable for oxide interfaces and surfaces. A group-theoretical decomposition revealed an effective spin-$\nicefrac{3}{2}$--orbit coupled Hamiltonian, resulting in additional quintet and septet channels in the Cooperon equation. The magneto-conductivity signature is determined by the topological spin winding number $N_\text{S}$ at the Fermi surface of the electronic system, causing an additional shoulder-like structure in the case $N_\text{S}=1$ and small triple winding contribution.

Previous fits for experimental data gained at LaAlO$_3$/SrTiO$_3$ interfaces identified a spin winding $N_\text{S}=3$ in the multi-band regime~\cite{PhysRevB.97.075136}. More recently, experiments near the Lifshitz transition in LaAlO$_3$/SrTiO$_3$ found indications for a dominant spin winding $N_\text{S}=1$ in the single-band regime, jointly with a considerable $N_\text{S}=3$ contribution beyond the Lifshitz transition in the multi-band regime~\cite{2019arXiv190403731Y}.
We expect that our results will help to identify the unique $N_\text{S}=1$ case in more multi-band situations. 

Funded by the
Deutsche Forschungsgemeinschaft (DFG, German Research Foundation) -- Grant number 107745057 -- TRR 80.
We are grateful to D.~Vollhardt and G.~Hammerl for helpful discussions.  

\FloatBarrier



%

\end{document}